\documentclass[aps,pra,twocolumn,10pt,superscriptaddress,nofootinbib,balancelastpage]{revtex4-1} 
\usepackage[latin1]{inputenc}
\usepackage{amsmath,amssymb}
\usepackage{mathrsfs} 
\usepackage[capitalise]{cleveref}
\usepackage{braket}
\usepackage{graphicx,color,colortbl}
\usepackage{booktabs}
\usepackage{dsfont}

\allowdisplaybreaks

\newcommand{\dif}{\mathrm{d}}%
\newcommand{\Nabla}{\vec{\nabla}}%
\newcommand{\Laplace}{\boldsymbol{\triangle}}%
\newcommand{\norm}[1]{\lVert#1\rVert}%
\newcommand{\ZT}[1]{\textquotedblleft#1\textquotedblright}%

\newlength{\myl}%
\newcommand{\INT}[3]{\settowidth{\myl}{$\displaystyle\int_{#1}^{#2}$}{\int_{#1}^{#2}\;\;\;\hspace{-\the\myl}\dif #3}\,}
\newcommand{\TINT}[3]{\settowidth{\myl}{$\displaystyle\int_{#1}^{#2}$}{\int_{#1}^{#2}\;\;\;\;\;\,\hspace{-\the\myl}\dif #3}\,}
\newcommand{\EINT}[3]{\settowidth{\myl}{$\int_{#1}^{#2}$}{\int_{#1}^{#2}\;\;\;\,\hspace{-\the\myl}\dif #3}\,}

\makeatletter
\newcommand\footnoteref[1]{\protected@xdef\@thefnmark{\ref{#1}}\@footnotemark}
\makeatother
\newcommand{\smallsim}{\smallsym{\mathrel}{\sim}}

\makeatletter
\newcommand{\smallsym}[2]{#1{\mathpalette\make@small@sym{#2}}}
\newcommand{\make@small@sym}[2]{%
  \vcenter{\hbox{$\m@th\downgrade@style#1#2$}}%
}
\newcommand{\downgrade@style}[1]{%
  \ifx#1\displaystyle\scriptstyle\else
    \ifx#1\textstyle\scriptstyle\else
      \scriptscriptstyle
  \fi\fi
}
\makeatother

\begin{document}

\title{Collective dynamics of active Brownian particles in three spatial dimensions: a predictive field theory}

\author{Jens Bickmann}
\affiliation{Institut f\"ur Theoretische Physik, Center for Soft Nanoscience, Westf\"alische Wilhelms-Universit\"at M\"unster, D-48149 M\"unster, Germany}

\author{Raphael Wittkowski}
\email[Corresponding author: ]{raphael.wittkowski@uni-muenster.de}
\affiliation{Institut f\"ur Theoretische Physik, Center for Soft Nanoscience, Westf\"alische Wilhelms-Universit\"at M\"unster, D-48149 M\"unster, Germany}

\begin{abstract}
We investigate the nonequilibrium dynamics of spherical active Brownian particles in three spatial dimensions that interact via a pair potential. The investigation is based on a predictive local field theory that is derived by a rigorous coarse-graining starting from the overdamped Langevin dynamics of the particles. This field theory is highly accurate and applicable even for the highest activities. It includes configurational order parameters and derivatives up to infinite orders. We present also three finite reduced models that result from the general field theory by suitable approximations and are easier to apply. Furthermore, we use the general field theory and the simplest one of the reduced models to derive analytic expressions for the density-dependent mean swimming speed and the spinodal corresponding to the onset of motility-induced phase separation of the particles, respectively. Both of these results show a good agreement with recent findings described in the literature. The analytic result for the spinodal yields also a prediction for the associated critical point whose position has not been determined before.  
\end{abstract}
\maketitle

\section{Introduction}
Active Brownian particles (ABPs), combining Brownian motion with self-propulsion, constitute a prime example for active matter that is currently attracting a lot of research interest \cite{Romanczuk2012,WensinkLMHWZKM2013,MIPS,Elgeti2015,BechingerEA16,Fodor16,Speck2016,Zttl2016,Marconi2017,Mallory18}. Besides artificial self-propelled microparticles \cite{Rao2015,Wu2016,Xu2016,Guix2018,Chang2019,PachecoJerez2019}, motile microorganisms \cite{SchwarzLinek2016,Chen2017,Andac2019} are often described as ABPs. It was shown that even bacteria, which perform a run-and-tumble motion \cite{Tailleur2008,Cates2013,Liu2017,Andac2019} like \textit{Escherichia coli} \cite{Berg2008,Tailleur2008,Paoluzzi_2013,Liang2018}, can successfully be described as ABPs.
Already isometric, i.e., spherical, ABPs have been found to exhibit many striking effects \cite{Elgeti2015,NegativeInterfaceTensionBialk15,MIPS,Ni15,Solon2015Nat,SolonSWKKCT2015,BechingerEA16,Takatori17,Duzgun2018,Tjhung2018,Das2019}.
Among them is motility-induced phase separation (MIPS) \cite{MIPS}, which became particularly popular and is still gaining great scientific interest \cite{Tailleur2008, Fily2012, Bialk2013, Buttinoni2013, Redner2013, Stenhammar2013, Speck2014, WittkowskiTSAMC2014, Wysocki2014, Zoettl2014, SolonSWKKCT2015, Redner2016, WittkowskiSC2017, Digregorio2018, Paliwal2018, Solon2018, Whitelam2018, Nie2019}. MIPS is the separation of the particles into a low-density gas-like and a high-density fluid-like phase that originates from the nonequilibrium behavior of the particles and occurs even if their interactions are purely repulsive.

The collective dynamics of ABPs and the arising effects can be described theoretically by field theories \cite{BickmannW2019}. This approach usually allows deeper insights into a system's behavior than computer simulations. As nonlocal field theories are typically much more difficult to interpret and treat numerically, most of the existing field theories for ABPs are local. Examples for nonlocal field theories describing ABPs are active dynamical density functional theories (DDFTs) \cite{Rex2007, WittkowskiL2011, Menzel2016j, teVrugtLW2020}. These theories, however, are applicable only for a weak propulsion of the particles and sometimes also only for low particle densities. Examples for local field theories include active phase field crystal (PFC) models \cite{EmmerichEtAl2012,Menzel2013,Menzel2014,Alaimo2016,Alaimo2018,PraetoriusVWL2018}. They are derived from DDFTs and thus include the same aforementioned limitations regarding the propulsion and density of the particles. Other local field theories include active diffusion equations \cite{Cates2013,Bialk2013}, an extension for mixtures of active and passive Brownian particles \cite{WittkowskiSC2017}, a hydrodynamic model explicitly considering the flow field of the particle suspension \cite{Steffenoni17}, a model with an explicit particle-field representation \cite{Grossmann19}, Cahn-Hilliard-like models \cite{Stenhammar2013, Speck2014, Solon2018}, the related nonintegrable Active Model B (AMB) \cite{WittkowskiTSAMC2014}, its extension Active Model B + (AMB+) \cite{Tjhung2018,cates_tjhung_2018}, and a recently published highly accurate predictive field theory \cite{BickmannW2019}.   
These models, however, are limited to weak self-propulsion, include only terms with low orders of the order-parameter fields and derivatives, or are restricted to systems with two spatial dimensions. Furthermore, the existing models are often not predictive. These strong limitations restrain the applicability or accuracy of the models. For example, the dimensionality of a system of ABPs is known to be crucial for its dynamical properties \cite{Stenhammar2014} so that a suitable theory for three spatial dimensions is needed \cite{BickmannW2019}. 

In this article, we present such a field theory. It is local and describes the nonequilibrium dynamics of interacting spherical ABPs in three spatial dimensions. The field theory is highly accurate and applicable even for the highest activities of the particles. For this purpose, approximations are kept to a minimum during the derivation of the theory. In its initial form, the theory takes configurational order-parameter fields and derivatives up to infinite order into account. Moreover, the theory is predictive, meaning that it includes analytic expressions that allow to calculate the values of all coefficients occurring in the field theory from the microscopic parameters of the system. This is achieved via a rigorous coarse-graining of the particles' dynamics that is described by overdamped Langevin equations. 
In addition to the general field theory, we provide three relatively simple reduced models of increasing complexity that involve only a small number of terms and are easier to apply than the general field theory. These models are obtained from the general theory by restricting the maximal order of order-parameter fields and derivatives as well as performing a quasi-stationary approximation. 
We discuss these models and show that several of the aforementioned local field theories for systems of ABPs from the literature can be identified as special cases of our reduced models. 
To demonstrate examples for concrete applications of our general field theory and reduced models, we use the general field theory to derive an analytic expression for the density-dependent mean swimming speed of ABPs in a homogeneous state. Besides the usually considered linear density dependence, our expression includes a nonlinear correction that is in excellent agreement with previous analytic findings described in Ref.\ \cite{Sharma2016}.  
Furthermore, we use the simplest of our reduced models to obtain an analytic expression for the spinodal corresponding to the onset of MIPS in a system of ABPs with an arbitrary pair-interaction potential. The obtained expression is found to be in good agreement with results of computer simulations presented in Ref.\ \cite{Broeker2019}. Our spinodal condition also yields a prediction for the associated critical point for which simulation results from the literature seem to be not yet available. 

The article is structured as follows: In section \ref{sec:derivation} we derive the general field theory and present possible simplifying approximations. The reduced models are presented and compared to other existing theories in section \ref{sec:MODELS}. In section \ref{sec:Applications}, exemplary applications of the field theory and reduced models are demonstrated. Finally, we conclude in section \ref{sec:Conclusion}.

\section{\label{sec:derivation}General field theory and approximations}
\subsection{\label{ssec:gfe}General field equations}
Our derivation of a general field theory is based on the \textit{interaction-expansion method} \cite{BickmannW2019}.
We consider $N$ similar active Brownian spheres with diameter $\sigma$ in three spatial dimensions. Their motion is described by their center-of-mass positions $\lbrace \vec{r}_i(t)\rbrace$ and orientational unit vectors $\lbrace \hat{u}_i(t) \rbrace$ as functions of time $t$, where the index $i\in\lbrace1,\dotsc,N \rbrace$ distinguishes between the particles. One can parameterize the vectors $\vec{r}_i = (x_{1, i}, x_{2, i}, x_{3, i})^{\mathrm{T}}$ and $\hat{u}_i = (\cos(\phi_i)\sin(\vartheta_i), \sin(\phi_i)\sin(\vartheta_i), \cos(\vartheta_i))^{\mathrm{T}}$ by the Cartesian coordinates $x_{1, i}$, $x_{2, i}$, and $x_{3, i}$ and the spherical coordinates $\vartheta_i$ and $\phi_i$, respectively. The equations of motion for the particles are given by the overdamped Langevin equations \cite{Stenhammar2014,Wysocki2014,Zttl2016,Paliwal2018,Broeker2019,Das2019}
\begin{align}
\dot{\vec{r}}_i &= \vec{\xi}_{\mathrm{T},i} + v_0\hat{u}_i+ \beta D_\mathrm{T} \vec{F}_{\mathrm{int}, i}, \label{eqn:LangevinR}\\
\dot{\hat{u}}_i &= \hat{u}_i\times\vec{\xi}_{\mathrm{R},i}. \label{eqn:LangevinPHI}%
\end{align}
Here, a dot over a variable denotes a derivative with respect to time, $v_0$ is the propulsion speed of a noninteracting particle, and $\beta =1/(k_{\mathrm{B}}T)$ is the thermodynamic beta with the Boltzmann constant $k_\mathrm{B}$ and the absolute temperature $T$ of the particles' environment. 
The translational diffusion constant of a particle is denoted as $D_{\mathrm{T}}$ and, since the particles are spherical, related to the rotational diffusion constant $D_{\mathrm{R}}$ by the expression $D_{\mathrm{R}}=3D_{\mathrm{T}}/\sigma^2$. 
With the statistically independent Gaussian white noises $\vec{\xi}_{\mathrm{T}, i}(t)$ and $\vec{\xi}_{\mathrm{R}, i}(t)$ the translational and rotational Brownian motion of the $i$-th particle is taken into account. 
Their correlations are given by $\braket{\vec{\xi}_{\mathrm{T}, i}(t_1)\otimes\vec{\xi}_{\mathrm{T}, j}(t_2)} = 2D_\mathrm{T}\delta_{ij}\mathds{1}_3\delta(t_1-t_2)$ and $\braket{\vec{\xi}_{\mathrm{R}, i}(t_1)\otimes\vec{\xi}_{\mathrm{R}, j}(t_2)} = 2D_\mathrm{R}\delta_{ij}\mathds{1}_3\delta(t_1-t_2)$ with the $3\times3$-dimensional identity matrix $\mathds{1}_3$.
Furthermore, $\vec{F}_{\mathrm{int},i}(\{\vec{r}_i\})$ describes the interaction force that the other $N-1$ particles exert on the $i$-th particle. 
We assume that the interactions can be described via a pair-interaction potential $U_2(\norm{\vec{r}_i-\vec{r}_j})$ so that the interaction force can be written as $\vec{F}_{\mathrm{int},i}(\{\vec{r}_i\}) = -\sum_{j=1, j\neq i}^{N}\Nabla_{\vec{r}_i} U_2(\norm{\vec{r}_i-\vec{r}_j})$, where $\Nabla_{\vec{r}_i} = (\partial_{x_{1, i}}, \partial_{x_{2, i}}, \partial_{x_{3, i}})^\mathrm{T}$ is the nabla operator that involves partial derivatives with respect to the elements of the position vector $\vec{r}_i$. 

Rewriting the Langevin equations \eqref{eqn:LangevinR} and \eqref{eqn:LangevinPHI} into the corresponding Smoluchowski equation yields 
\begin{equation} 
\begin{split}
\dot{\mathfrak{P}} & = \sum_{i=1}^N \big((D_\mathrm{T}\Laplace_{\vec{r}_i} + D_\mathrm{R}\mathcal{R}_i^2 )\mathfrak{P}\\
&\qquad\quad\,\:\!-\Nabla_{\vec{r}_i}\cdot \big( (v_0\hat{u}_i + \beta D_\mathrm{T} \vec{F}_{\mathrm{int}, i}) \mathfrak{P}\big)\big)
\end{split}
\end{equation}
with the many-particle probability density $\mathfrak{P}(\{\vec{r}_i\},\{\hat{u}_i\},t)$. The symbol $\Laplace_{\vec{r}_i}=\Nabla_{\vec{r}_i}\cdot\Nabla_{\vec{r}_i}$ denotes the Laplacian and $\mathcal{R}_i = \hat{u}_i\times\Nabla_{\hat{u}_i}$ is the rotational operator with $\Nabla_{\hat{u}_i}$ being the nabla operator that involves partial derivatives with respect to the elements of the orientation vector $\hat{u}_i$.
By integrating over all degrees of freedom except for those of one particle and renaming its position as $\vec{r}$ and its orientation as $\hat{u}$, one gets a dynamic equation for the local orientation-dependent density
\begin{align}
\varrho(\vec{r}, \hat{u}, t) = N \bigg(\prod_{\begin{subarray}{c}j = 1\\j\neq i\end{subarray}}^{N} \int_{\mathbb{R}^3}\!\!\!\!\!\dif^{3}r_j\int_{\mathbb{S}_2}\!\!\!\!\dif^{2} u_j \bigg)\, \mathfrak{P}\bigg\rvert_{\begin{subarray}{l}\vec{r}_i=\vec{r},\\\hat{u}_i=\hat{u}\end{subarray}}.
\label{eqn:OPD}%
\end{align}
The symbol $\mathbb{S}_2$ denotes the three-dimensional unit sphere and $\dif^{2} u_j=\sin(\vartheta_j)\dif\vartheta_j\dif\phi_j$ is a solid angle differential. Using the divergence theorem and neglecting boundary terms, one obtains 
\begin{equation}
\begin{split}%
\dot{\varrho} &= ( D_\mathrm{T}\Laplace_{\vec{r}}+D_\mathrm{R}\mathcal{R}^2- v_0\Nabla_{\vec{r}}\cdot\hat{u}) \varrho + \mathcal{I}_\mathrm{int} \label{eqn:NeededforV}%
\end{split}%
\end{equation}
with the interaction term
\begin{equation}
\begin{split}
\mathcal{I}_\mathrm{int} &= \beta D_\mathrm{T}\Nabla_{\vec{r}}\cdot \bigg( \varrho(\vec{r}, \hat{u}, t)\int_{\mathbb{R}^3}\!\!\!\!\!\dif^3r'\, U_2'(\norm{\vec{r}-\vec{r}'})\\
&\quad\,\:\! \frac{\vec{r}-\vec{r}'}{\norm{\vec{r} - \vec{r}'}} \int_{\mathbb{S}_2}\!\!\!\!\dif^{2}u'\, g(\vec{r}, \vec{r}', \hat{u}, \hat{u}', t)
\varrho(\vec{r}', \hat{u}', t) \bigg),
\end{split}
\label{eqn:Iint}%
\end{equation}
where $U_2'(r) \equiv \dif U_2(r)/\dif r$ is a shorthand notation. The pair-distribution function $g(\vec{r}, \vec{r}', \hat{u}, \hat{u}', t)$ describes the relation between the two-particle density $\varrho^{(2)}(\vec{r}, \vec{r}', \hat{u}, \hat{u}', t)$ and the one-particle density product $\varrho(\vec{r}, \hat{u}, t)\varrho(\vec{r}', \hat{u}', t)$:
\begin{equation}
\varrho^{(2)}(\vec{r}, \vec{r}', \hat{u}, \hat{u}', t) = g(\vec{r}, \vec{r}', \hat{u}, \hat{u}', t)\varrho(\vec{r}, \hat{u}, t)\varrho(\vec{r}', \hat{u}', t).
\end{equation}

\begin{figure}[t]
\centering
\includegraphics{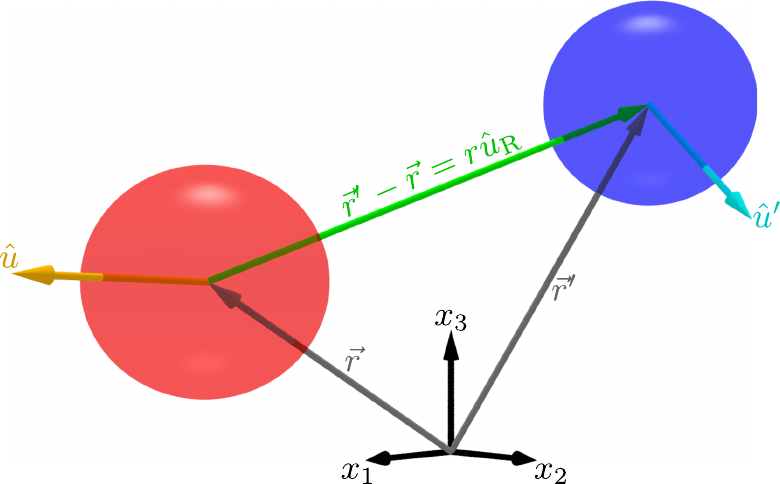}%
\caption{\label{fig:geometry}Notation for the positions and orientations of two active Brownian spheres in three spatial dimensions.}
\end{figure}

We assume the system to be in a homogeneous stationary state so that the pair-distribution function can be parameterized as $g(r, \hat{u}_{\mathrm{R}}, \hat{u}, \hat{u}')$, where the particle distance $r$ and the unit vector $\hat{u}_{\mathrm{R}}$ are defined by the relation $\vec{r}'-\vec{r} \equiv r\hat{u}_{\mathrm{R}}$. A sketch of the geometry is shown in Fig.\ \ref{fig:geometry}.

Taking the rotational invariance of the homogeneous state into account, we expand the pair-distribution function into spherical harmonics $Y_{l}^{m}(\hat{u})$ \cite{GubbinsGray}:\footnote{We use the Condon-Shortley phase convention for the spherical harmonics \cite{GubbinsGray}.} 
\begin{equation}
\begin{split}
g(r,\hat{u}_{\mathrm{R}}, \hat{u}, \hat{u}') &= \sum_{l_rll'} g(l_rll'; r)\!\!\!\sum_{m_rmm'}\!\!\!C(ll'l_r, mm'm_r)\\
&\qquad\;\;\; Y_{l_r}^{m_r*}(\hat{u}_{\mathrm{R}})Y_{l}^{m}(\hat{u})Y_{l'}^{m'}(\hat{u}').
\end{split}\label{eq:sh_expansion}\raisetag{2.6em}
\end{equation}
Here, 
\begin{equation}
\sum_{l_rll'} \sum_{m_rmm'} =\sum_{l_r,l,l' = 0}^{\infty} \sum_{m_r = -l_r}^{l_r}\sum_{m = -l}^{l}\sum_{m' = -l'}^{l'}
\end{equation}
is a shorthand notation, 
\begin{widetext}
\begin{equation}
g(l_rll'; r) = \begin{cases}
\frac{\sqrt{(2l_r+1)(2l+1)(2l'+1)}}{(4\pi)^{3/2} C(ll'l_r, 000)} \int_{\mathbb{S}_2}\!\!\!\dif^{2}u\int_{\mathbb{S}_2}\!\!\!\dif^{2}u'\int_{\mathbb{S}_2}\!\!\!\dif^{2}u''\, g(r, \hat{u}, \hat{u}', \hat{u}'') P_{l_r}(\hat{u})P_{l}(\hat{u}')P_{l'}(\hat{u}''), &\text{if } l+l'+l_r=\text{even,}\\
0, &\text{else}
\end{cases}
\end{equation}
\end{widetext} 
are the expansion coefficients of the pair-distribution function $g(r,\hat{u}_{\mathrm{R}}, \hat{u}, \hat{u}')$ with the Legendre polynomials $P_l(\hat{u})$, $C(ll'l_r, mm'm_r)$ are the Clebsch-Gordan coefficients, and the superscript $^*$ denotes complex conjugation. 

Similar to the spherical harmonics expansion of the pair-distribution function, an orientational expansion of the one-particle density field $\varrho(\vec{r}, \hat{u},t)$ into Cartesian tensor order-parameter fields $\mathcal{O}_{i_1 \dotsb i_n}(\vec{r}, t)$  \cite{GubbinsGray,teVrugtW2019b,BickmannW2019} is performed:\footnote{From here onwards, summation over indices appearing twice in a term is implied.}  
\begin{equation}
\varrho(\vec{r}, \hat{u}, t) = \sum_{n=0}^{\infty} \mathcal{O}_{i_1\dotsb i_n}(\vec{r}, t) u_{i_1}\!\dotsb u_{i_n}.
\label{eqn:ODEx2}%
\end{equation}
The Cartesian order-parameter fields $\mathcal{O}_{i_1 \dotsb i_n}(\vec{r}, t)$ can be obtained as 
\begin{equation}
\mathcal{O}_{i_{1}\dotsb i_{n}}(\vec{r}, t) = \int_{\mathbb{S}_{2}}\!\!\!\!\dif^{2}u\, \mathfrak{U}_{i_{1}\dotsb i_{n}}(\hat{u}) \varrho(\vec{r},\hat{u},t), 
\label{eqn:ODEx}%
\end{equation}
with the orientation-dependent $n$-th-order tensors $\mathfrak{U}_{i_1\dotsb i_n}(\hat{u})$. These tensors are defined by the expansion \eqref{eqn:ODEx2} \cite{teVrugtW2019b}.
The elements of the orientation vector $\hat{u}$ are denoted as $u_i = (\hat{u})_i$. 

Applying the aforementioned expansions as well as an untruncated gradient expansion in the interaction term \eqref{eqn:Iint}, the integrations in this term can be carried out and a local expression is obtained. Inserting this expression into Eq.\ \eqref{eqn:NeededforV} and using Eq.\ \eqref{eqn:ODEx} leads to the following dynamic equation for the configurational order-parameter fields $\mathcal{O}_{i_1 \dotsb i_n}(\vec{r}, t)$:
\begin{widetext}
\begin{equation}
\begin{split}
\dot{\mathcal{O}}_{i_1 \dotsb i_d} =&\, \sum_{k=0}^{\infty}\int_{\mathbb{S}_2}\!\!\!\!\dif^{2}u\, \mathfrak{U}_{i_1 \dotsb i_d} \Bigg((D_\mathrm{T}\Laplace_{\vec{r}} + D_\mathrm{R}\mathcal{R}^2 - v_0\partial_bu_b) u_{j_1}\!\dotsb u_{j_k} \mathcal{O}_{j_1\dotsb j_k} \\
&\qquad\qquad\qquad\qquad + \frac{4}{\sqrt{\pi}}\partial_{p_0}\Bigg(
\sum_{h=0}^{\infty} u_{a_1}\!\dotsb u_{a_h} \mathcal{O}_{a_1\dotsb a_h}
\sum_{n=0}^{\infty}\sum_{l_rll'}\sum_{m_rmm'}C(ll'l_r, mm'm_r) Y_{l}^{m}(\hat{u}) \\
&\qquad\qquad\qquad\qquad\qquad\qquad\qquad\;\;\,\:\!
\dfrac{1}{n!} G(n, l_r, l, l') (\mathcal{S}_{l_r}^{m_r})_{p_0p_1 \dotsb p_n}(\mathcal{S}_{l'}^{m'})_{j_1 \dotsb j_k} \nabla_{p_1 \dotsb p_{n}}\mathcal{O}_{j_1\dotsb j_k}
\Bigg)\!\Bigg).
\end{split}
\label{eqn:formalDynEqOP}%
\end{equation} 
\end{widetext}
For a compact notation, we introduce the spherical coefficients
\begin{equation}
G(n, l_r,l,l') = -\frac{\sqrt{\pi}}{4}\beta D_\mathrm{T}\int_{0}^{\infty}\!\!\!\!\!\!\dif r'\, r'^{n+2}U_2'(r') g(l_r l l'; r'),
\label{eqn:CoeffDef}%
\end{equation}
the spherical coefficient tensors
\begin{equation}
(\mathcal{S}_{l}^{m})_{i_1 \dotsb i_n} = \int_{\mathbb{S}_2}\!\!\!\dif^{2}u\, Y_{l}^{m}(\hat{u}) u_{i_1}\!\dotsb u_{i_n}
\end{equation}
with the property $(\mathcal{S}_{l}^{m})_{i_1 \dotsb i_N} = 0\quad \forall\, l>N$, and the operators $\nabla_{i_1 \dotsb i_m} \equiv \partial_{i_1}\!\dotsb\partial_{i_m}$, where $\partial_i$ is the $i$-th component of the nabla operator acting on $\vec{r}$. The field equations \eqref{eqn:formalDynEqOP} describe the time evolution of the order-parameter fields $\mathcal{O}_{i_1 \dotsb i_d}$ and constitute a rather general field theory that is the main result of the present work.

Our field equations \eqref{eqn:formalDynEqOP} extend those from Ref.\ \cite{BickmannW2019} to three spatial dimensions. By restricting both the positions and orientations of the particles to a plane, the field theory \eqref{eqn:formalDynEqOP} simplifies to that of Ref.\ \cite{BickmannW2019}. The spherical harmonics expansion \eqref{eq:sh_expansion} reduces to a Fourier expansion and the Clebsch-Gordan coefficients $C(ll'l_r, mm'm_r)$ reduce to Kronecker symbols $\delta_{l_r+l', l} = \delta_{k_3, k_1+k_2}$. Thus it is possible to relate the coefficients $A(n,k_1,k_2)$ from Ref.\ \cite{BickmannW2019} to the coefficients $G(n,l_r,l,l')$ of the present theory:
\begin{equation}
G(n, a, b, c) \smallsim A(n, b+c, b-a).
\label{eqn:CoeffRelation}%
\end{equation}
To calculate the values of the coefficients \eqref{eqn:CoeffDef}, the pair-distribution function $g(r,\hat{u}_{\mathrm{R}},\hat{u}, \hat{u}')$ has to be, at least partially, known. 
Results for the pair-distribution function can be obtained by analytic approaches and computer simulations. For ABPs interacting by a Weeks-Chandler-Andersen potential, there exist analytic representations for two \cite{JeggleSW2019} and three \cite{Broeker2019} spatial dimensions that can be used.
The analytic representation from Ref.\ \cite{Broeker2019} will be used in section \ref{sec:Applications} further below.

\subsection{\label{ssec:APPROXIMATIONS}Approximations}
The general field equations \eqref{eqn:formalDynEqOP} are rather complicated, since they contain order-parameter fields and derivatives up to order infinity. However, the field equations can be seen as a framework from which simpler reduced models can be obtained by truncating the maximal order of order-parameter fields and derivatives.

The first three order-parameter fields are the density field $\rho(\vec{r}, t)$ describing the local particle number density, the polarization vector $\vec{\mathcal{P}}(\vec{r}, t)$ that gives insights into the local mean orientation of the particles, and the symmetric and traceless nematic tensor $\mathcal{Q}_{ij}(\vec{r}, t)$ that characterizes their local alignment. Restricting the orientational expansion of the one-particle density to these three order parameters is common both in theories for liquid crystals \cite{deGennesP1995, EmmerichEtAl2012, WittkowskiLB2010, WittkowskiLB2011, WittkowskiLB2011b, PraetoriusVWL2013} and active matter \cite{Menzel2013, Menzel2014, TiribocchiWMC2015, StenhammarWMC2016, WittkowskiSC2017, PraetoriusVWL2018, BickmannW2019}. 
This means that the one-particle density $\varrho(\vec{r}, \hat{u} ,t)$ is approximated as
\begin{equation}
\varrho(\vec{r},\hat{u},t) \approx \frac{\rho(\vec{r}, t)}{4\pi} + \mathcal{P}_i(\vec{r}, t)u_i + \mathcal{Q}_{ij}(\vec{r}, t)u_iu_j.
\end{equation}
We will use this approximation, \textit{truncating the orientational expansion} of the one-particle density, throughout the rest of this work. 
The order-parameter fields can be obtained from the one-particle density via their definitions
\begin{align}
\rho(\vec{r}, t) =&  \int_{\mathbb{S}_2}\!\!\!\dif^{2}u\,  \varrho(\vec{r}, \hat{u}, t), \label{eqn:projRho} \\
\mathcal{P}_i(\vec{r}, t) =& \frac{3}{4\pi} \int_{\mathbb{S}_2}\!\!\!\dif^{2}u\, \varrho(\vec{r}, \hat{u}, t) u_i, \label{eqn:projP}\\
\mathcal{Q}_{ij}(\vec{r}, t) =& \frac{15}{8\pi} \int_{\mathbb{S}_2}\!\!\!\dif^{2}u\, \varrho(\vec{r}, \hat{u}, t) \Big(u_iu_j -\frac{1}{3}\delta_{ij}\Big),\label{eqn:projQ}
\end{align}
which allow for the identification of the corresponding tensors $\mathfrak{U}_{i_1\dotsb i_n}(\hat{u})$.

When \textit{truncating the gradient expansion}, a maximal order of two, four, or six is typically chosen \cite{BickmannW2019}.  
A model containing a maximal order of two derivatives can describe the enhanced mobility of ABPs and the onset of instabilities, allowing, e.g., to calculate the spinodal for MIPS \cite{Bialk2013,WittkowskiSC2017, Nie2019, BickmannW2019}. If a further analysis of the system of ABPs is desired, $4$th-order-derivatives models, often called phase field model, can be used to describe also the emergence of patterns and their dynamics \cite{Speck2014,WittkowskiTSAMC2014,Tjhung2018}. A $6$th-order-derivatives model can, on top of that, describe crystallization and particle-resolved lattice structures \cite{Menzel2013,Menzel2014, Alaimo2016,Alaimo2018,PraetoriusVWL2018}. Models of odd orders in derivatives are typically not considered \cite{BickmannW2019}. 

Considering the aforementioned restrictions for the order in the orientational and gradient expansions yields models with three coupled time-dependent partial differential equations for $\rho$, $\mathcal{P}_i$, and $\mathcal{Q}_{ij}$. A further reduction of the complexity of the models can be achieved by performing a \textit{quasi-stationary approximation} (QSA) \cite{Cates2013,WittkowskiSC2017, BickmannW2019}. This approximation exploits the fact that, unlike the density $\rho$, the polarization vector $\mathcal{P}_i$ and the nematic tensor $\mathcal{Q}_{ij}$ are nonconserved quantities whose dynamics can be considered as instantaneously relaxing when the system is described on the characteristic time scale of the density. A QSA results in a dynamic equation for $\rho$ and constitutive equations for $\mathcal{P}_i$ and $\mathcal{Q}_{ij}$, where all three equations include only the order-parameter field $\rho$. Applying a QSA conserves the locality of a model and usually the maximal order of derivatives is kept unchanged in the model's equations. 

Lastly, if a model is still considered to be too complex, a \textit{low-density approximation} \cite{BickmannW2019} can be applied to further reduce the number of terms in the model. This approximation truncates the total order with respect to the density and derivatives of the individual terms at a chosen cutoff value. 

We point out that these approximations, especially the QSA and the low-density approximation, are not necessary and can be omitted if desired.

\section{\label{sec:MODELS}Reduced models}
In this section, we present three reduced models obtained by making use of the approximations described in section \ref{ssec:APPROXIMATIONS}. These models are a $2$nd-order-derivatives model, a $4$th-order-derivatives model, and a $7$th-order low-density model. All of these models were obtained via a QSA. The last one involves also a low-density approximation. The models consist of a continuity equation
\begin{equation}
\dot{\rho} = \partial_i J_i
\label{eqn:ConsservativeForm}%
\end{equation}
for the density with the density current $J_i$ and constitutive equations for $\mathcal{P}_i$ and $\mathcal{Q}_{ij}$. In the equations for $\mathcal{P}_i$ and $\mathcal{Q}_{ij}$, we omit the highest one and two orders, respectively, since they do not contribute to the density current $J_i$ \cite{BickmannW2019}. We compare our models with popular models for ABPs in three spatial dimensions from the literature.\footnote{The models from the literature can be applicable also to systems that are not considered in the present work. We neglect the corresponding additional features of these models when comparing with our models.} 
The models from Refs.\ \cite{Steffenoni17} and \cite{Grossmann19} are not included in our comparison, since they are ABP models in two spatial dimensions.

\subsection{\label{sec:2ndorderGradientModel}$\boldsymbol{2}$nd-order-derivatives model}
The first reduced model contains derivatives up to second order. Its density current is given by
\begin{equation}
J_i = D(\rho)\partial_i\rho
\label{eqn:Dyn2DwD}%
\end{equation}
with the density-dependent diffusion coefficient
\begin{equation}
\begin{split}
D(\rho) &= D_\mathrm{T}+\frac{2}{3\pi}G(1,0,0,0)\rho\\
&\quad\, +\frac{1}{6D_\mathrm{R}}\Big(v_0-\frac{4}{\pi}G(0,1,1,0)\rho\Big)\\
&\qquad\qquad\;\;\,\Big(v_0-\frac{2}{\pi}(G(0,1,1,0)+G(0,1,0,1))\rho\Big).
\end{split}\raisetag{4.2em}
\label{eqn:diffcoeffienetn}%
\end{equation}
The constitutive equations for the polarization vector and nematic tensor read
\begin{equation}
\mathcal{P}_i = -\frac{v_0-\frac{4}{\pi}G(0,1,1,0)\rho}{8\pi D_\mathrm{R}}\partial_i\rho
\end{equation}
and
\begin{equation}
\mathcal{Q}_{ij} = 0,
\label{eqn:Qin2nd}%
\end{equation}
respectively. This is the model of the lowest nontrivial order in derivatives that can be obtained from the general field theory \eqref{eqn:formalDynEqOP}.

The form of Eqs.\ \eqref{eqn:Dyn2DwD}-\eqref{eqn:Qin2nd} is similar to that of the corresponding model for two spatial dimensions, which is given by Eqs.\ (20)-(24) in Ref.\ \cite{BickmannW2019}. 
This means that, within these models, the dynamic behavior of ABPs in two and three spatial dimensions is similar. 
Three different coefficients $G(1, 0, 0, 0)$, $G(0,1,1,0)$, and $G(0,1,0,1)$ contribute to the density-dependent diffusion coefficient \eqref{eqn:diffcoeffienetn}. 
Neglecting the coefficients $G(1, 0, 0, 0)$ and $G(0,1,0,1)$ reduces the diffusion coefficient \eqref{eqn:diffcoeffienetn} to that given by Eq.\ (10) in Ref.\ \cite{Cates2013}, when selecting a dimensionality $d=3$ and run-and-tumble rate $\alpha = 0$. 
A comparison of our model with that of Ref.\ \cite{Cates2013} also yields the expression 
$v(\rho)=v_0-\frac{2}{\pi}G(0,1,1,0)\rho$ for the phenomenological density-dependent swimming speed in the latter model.

\subsection{$\boldsymbol{4}$th-order-derivatives model}
The second model that we present contains up to four derivatives per term. Its density current reads
\begin{equation}
\begin{split}
J_i &= (\alpha_{1}+\alpha_{2}\rho+\alpha_{3}\rho^2)\partial_i\rho\\
&\quad +(\alpha_{4}+\alpha_{5}\rho+\alpha_{6}\rho^2+\alpha_{7}\rho^3+\alpha_{8}\rho^4)\partial_i\Laplace\rho\\
&\quad +(\alpha_{9}+\alpha_{10}\rho+\alpha_{11}\rho^2+\alpha_{12}\rho^3)(\Laplace\rho)\partial_i\rho\\
&\quad +(\alpha_{13}+\alpha_{14}\rho+\alpha_{15}\rho^2+\alpha_{16}\rho^3)(\partial_i\partial_j\rho)\partial_j\rho\\
&\quad +(\alpha_{17}+\alpha_{18}\rho+\alpha_{19}\rho^2)(\partial_j\rho)(\partial_j\rho)\partial_i\rho 
\end{split}\label{eqn:4thorderCurrentJ}%
\end{equation}
and the constitutive equations for $\mathcal{P}_i$ and $\mathcal{Q}_{ij}$ are given by
\begin{equation}
\begin{split}
\mathcal{P}_i &= (\beta_{1}+\beta_{2}\rho)\partial_i\rho\\
&\quad +(\beta_{3}+\beta_{4}\rho+\beta_{5}\rho^2+\beta_{6}\rho^3)\partial_i\Laplace\rho\\
&\quad +(\beta_{7}+\beta_{8}\rho+\beta_{9}\rho^2)(\Laplace\rho)\partial_i\rho\\
&\quad +(\beta_{10}+\beta_{11}\rho+\beta_{12}\rho^2)(\partial_i\partial_j\rho)\partial_j\rho\\
&\quad +(\beta_{13}+\beta_{14}\rho)(\partial_j\rho)(\partial_j\rho)\partial_i\rho
\end{split}\label{eqn:P4th}%
\end{equation}
and
\begin{equation}
\begin{split}
\mathcal{Q}_{ij} &= (\gamma_1+\gamma_2\rho+\gamma_3\rho^2)(3\partial_i\partial_j\rho+\delta_{ij}\Laplace\rho)\\
&\quad +(\gamma_2+2\gamma_3\rho)(-3(\partial_i\rho)\partial_j\rho+\delta_{ij}(\partial_k\rho)\partial_k\rho),
\end{split}\label{eqn:Q4th}%
\end{equation}
respectively. Expressions for the coefficients $\alpha_1,\dotsc,\alpha_{19}$, $\beta_1,\dotsc,\beta_{14}$, $\gamma_1$, $\gamma_2$, and $\gamma_3$ can be found in the Appendix. According to its form, this model can also be called a \ZT{phase field model}. Neglecting all coefficients except for $\alpha_{1}$, $\alpha_{2}$, $\alpha_{3}$, $\beta_{1}$, and $\beta_{2}$ would result in the $2$nd-order-derivatives model. The $4$th-order-derivatives model is the first reduced model considered here, where the constitutive equation for the nematic tensor $\mathcal{Q}_{ij}$ does not vanish and it contributes to the dynamics of the concentration field. Again, the model has a similar form as the corresponding model for two spatial dimensions, which is given by Eqs.\ (20) and (25)-(27) in Ref.\ \cite{BickmannW2019}.

We compare our $4$th-order-derivatives model, which is given by Eqs.\ \eqref{eqn:Dyn2DwD} and \eqref{eqn:4thorderCurrentJ}-\eqref{eqn:Q4th}, with the models of Refs.\ \cite{Stenhammar2013, WittkowskiTSAMC2014, Solon2018, Tjhung2018}. All these models from the literature consist only of a dynamic equation for the density field and do not include equations for other order-parameter fields.

First, we compare our model with the general model of Ref.\ \cite{Solon2018}, which is given by a continuity equation for the density and the density current (2) in that work.\footnote{The authors of Ref.\ \cite{Solon2018} present their dynamic equation in two equivalent forms, which are given by Eqs.\ (1) and (2) in that work, and give relations between the coefficients of these different forms. For convenience, we compare with the second of these forms.} Their model is an out-of-equilibrium generalization of the Cahn-Hilliard model and allows its coefficients to be density-dependent. A comparison of their model with ours leads to the relations of the coefficients of both models that are listed in table \ref{tab:CoefficientComparisonSOLON}. 
The comparison shows that their model includes ours as a special case and leads to expressions for their general coefficients in terms of the density field and the predictive coefficients of our model. These relations allow to combine the framework of general thermodynamic variables of Ref.\ \cite{Solon2018} with our predictive model and thus to calculate, e.g., the binodal for a system of ABPs in three spatial dimensions \cite{Solon2018}. 
\begin{table}[ht]
\caption{\label{tab:CoefficientComparisonSOLON}Relations of the coefficients $\alpha(\rho)$, $\kappa(\rho)$, $\lambda(\rho)$, $\beta(\rho)$, and $\zeta(\rho)$ from the model given by a continuity equation for the density with the density current (2) in Ref.\ \cite{Solon2018} and the coefficients $\alpha_{1},\dotsc,\alpha_{19}$, $\beta_{1},\dotsc,\beta_{14}$, $\gamma_1$, $\gamma_2$, and $\gamma_3$ from the $4$th-order-derivatives model (or \ZT{phase field model}) given by Eqs.\ \eqref{eqn:ConsservativeForm} and \eqref{eqn:4thorderCurrentJ}-\eqref{eqn:Q4th} in the present work.}
\renewcommand{\arraystretch}{1.5}%
\begin{tabular}{|c|c|}
\hline
Relations of the coefficients & \begin{tabular}[c]{@{}c@{}}Corresp.\ term in the\\[-5pt]model from Ref.\ \cite{Solon2018}\end{tabular} \\
\hline
$\alpha(\rho) = -\alpha_{1}-\alpha_{2}\rho - \alpha_{3}\rho^2$ & $\partial_i\rho$\\
\begin{tabular}[c]{@{}l@{}}
$\kappa(\rho) = \alpha_{4}+\alpha_{5}\rho+\alpha_{6}\rho^2$ \\[-5pt]
$\qquad\,\,\,\,\;\; +\,\alpha_{7}\rho^3+\alpha_{8}\rho^4$\end{tabular}& $\partial_i\Laplace\rho$\\
$\zeta(\rho) = -\alpha_{9}-\alpha_{10}\rho-\alpha_{11}\rho^2-\alpha_{12}\rho^3 $& $(\Laplace\rho)\partial_i\rho$\\
\begin{tabular}[c]{@{}l@{}}
$\lambda(\rho) = -\frac{1}{2}(\alpha_{13}+\alpha_{14}\rho+\alpha_{15}\rho^2$\\[-5pt]
$\qquad\qquad\,\,\;\;+\,\alpha_{16}\rho^3)$\end{tabular} & $(\partial_i\partial_j\rho)\partial_j\rho$\\
$ \beta(\rho) = -\alpha_{17}-\alpha_{18}\rho-\alpha_{19}\rho^2$& $(\partial_j\rho)(\partial_j\rho)\partial_i\rho$\\
$\beta_{1},\dotsc,\beta_{14}=0$ & -- \\
$\gamma_1,\gamma_2,\gamma_3=0$ & -- \\[1pt]
\hline
\end{tabular}
\end{table}

The second model we compare with is the continuum theory given by Eqs.\ (10)-(13) in Ref.\ \cite{Stenhammar2013}. We neglect the phenomenological term $\mu_{\mathrm{rep}}$ of this model, which was included to take excluded-volume interactions into account, since it gives a contribution $\propto \rho^5\partial_i\rho$ in the density current that is not allowed by the combination of gradient expansion and QSA employed in the derivation of our model. 
We also neglect the noise vector $\vec{\Lambda}$ of the model of Ref.\ \cite{Stenhammar2013}, since our model is deterministic. 
The model of Ref.\ \cite{Stenhammar2013} can then be identified as a limiting case of our model, which is obtained when performing a nondimensionalization of our model and assuming the relations that are given in table \ref{tab:CoefficientComparison37}.
\begin{table}[htb]
\caption{\label{tab:CoefficientComparison37}The same as in table \ref{tab:CoefficientComparisonSOLON}, but now for the model given by Eqs.\ (10)-(13) in Ref.\ \cite{Stenhammar2013}, where $\kappa_0$ is a parameter of the model and the characteristic length $q_0$, characteristic time $t_0$, and reference density $\rho_0$ stem from the nondimensionalization underlying this model.}
\renewcommand{\arraystretch}{1.5}%
\begin{tabular}{|c|c|}
\hline
Relations of the coefficients & \begin{tabular}[c]{@{}c@{}}Corresp.\ term in the\\[-5pt]model from Ref.\ \cite{Stenhammar2013}\end{tabular} \\
\hline
$1=\frac{{t_0}}{{q_0^2}}\alpha_1 = -\frac{{t_0}}{3{q_0^2}}\rho_{0}  \alpha_2 = \frac{{t_0}}{2{q_0^2}}\rho_{0}^2\alpha_3$ & $\partial_i\rho$\\
\begin{tabular}[c]{@{}l@{}}$\kappa_0 = -\frac{{t_0}}{{q_0^4}}\rho_{0} \alpha_5 = \frac{{t_0}}{3{q_0^4}}\rho_{0}^2 \alpha_6 = -\frac{{t_0}}{3{q_0^4}}\rho_{0}^3 \alpha_7$ \\ \hskip2.9ex $=\frac{{t_0}}{{q_0^4}}\rho_{0}^4 \alpha_8$\end{tabular} & $\rho\partial_i\Laplace\rho$\\
$\kappa_0 = \frac{{t_0}}{{q_0^4}}\rho_{0}^{{2}} \alpha_{10} = -\frac{{t_0}}{2{q_0^4}}\rho_{0}^{{3}} \alpha_{11} = \frac{{t_0}}{{q_0^4}}\rho_{0}^{{4}} \alpha_{12} $ & $(\Laplace\rho)\partial_i\rho$\\
$\alpha_4,\alpha_9,\alpha_{13},\dots,\alpha_{19}=0$ & -- \\
$\beta_1,\dotsc,\beta_{14}=0$ & -- \\
$\gamma_1,\gamma_2,\gamma_3=0$ & -- \\[1pt]
\hline
\end{tabular}
\end{table}

Third, we compare our model with the models AMB and AMB+ presented in Refs.\ \cite{WittkowskiTSAMC2014,Tjhung2018}. Both of these models can be identified as limiting cases of our model. AMB+, which is given by Eqs.\ (3), (5), and (6) in Ref.\ \cite{Tjhung2018}, can be obtained from our model by performing a nondimensionalization and assuming the relations given in table \ref{tab:CoefficientComparison}.
The nondimensionalization introduces a dimensionless density field $\phi = c_\phi(\rho-\bar{\rho})$, where the constant $c_\phi$ accounts for the nondimensionality of $\phi$ and $\bar{\rho}$ is a reference density. In AMB and AMB+, $\bar{\rho}$ is chosen to be the mean-field density corresponding to the critical point \cite{WittkowskiTSAMC2014,Tjhung2018}. 
To keep AMB+ simple, the mobility $M$ in this model is considered to be independent of the density in Ref.\ \cite{Tjhung2018}, but, as mentioned in that work, the mobility could in principle depend on $\rho$ and even its derivatives. 
Our model can be written with such a mobility as well, if one considers a tensorial mobility.
Furthermore, AMB, which is given by Eqs.\ (1)-(3) in Ref.\ \cite{WittkowskiTSAMC2014}, is a limiting case of the more general AMB+ and therefore also included in our model as a limiting case. AMB is obtained from AMB+ when setting $\zeta=0$ and therefore from our model when performing a nondimensionalization, assuming the relations given in table \ref{tab:CoefficientComparison}, and additionally setting $\alpha_{9}=\alpha_{5}$.
\begin{table}[htb]
\caption{\label{tab:CoefficientComparison}The same as in table \ref{tab:CoefficientComparisonSOLON}, but now for the model AMB+ given by Eqs.\ (3), (5), and (6) in Ref.\ \cite{Tjhung2018}, where $a$, $b$, $K_0$, $K_1$, $\zeta$, $\lambda$, $\bar{\rho}$, $M$, and $c_\phi$ are parameters of the model and the characteristic length $q_0$ and characteristic time $t_0$ stem from the nondimensionalization underlying this model.}
\renewcommand{\arraystretch}{1.5}%
\begin{tabular}{|c|c|}
\hline
\;Relations of the coefficients\; & \begin{tabular}[c]{@{}c@{}}\;Corresp.\ term in the\;\\[-5pt]model from Ref.\ \cite{Tjhung2018}\end{tabular} \\[1.0pt]
\hline
$a = {t_0} \frac{\alpha_1-\bar{\rho}^2 \alpha_{{3}}}{{q_0^2} M}$ & $\partial_i\rho$\\
$b = \frac{{t_0} \alpha_{{3}}}{3 {q_0^2} c_\phi^2 M}$, $\alpha_2+2\bar{\rho}\alpha_3 = 0$ & $\partial_i(\rho^3)$\\
$K_0 = -{t_0}\frac{\alpha_4+\bar{\rho}\alpha_5}{{q_0^4} M}$& $\partial_i\Laplace\rho$\\
$K_1 = -\frac{{t_0} \alpha_5}{2 {q_0^4} c_\phi M}$ & $\rho\partial_i\Laplace\rho$, $(\partial_i\partial_j\rho)\partial_j\rho$\\
$\zeta =\frac{t_0}{{q_0^4} c_\phi M}(\alpha_5 -\alpha_{9})$ & $(\Laplace\rho)\partial_i\rho$ \\
$\lambda = {t_0} \frac{\alpha_{13}-2\alpha_5}{2 {q_0^4} c_\phi M}$& $(\partial_i\partial_j\rho)\partial_j\rho$ \\
\begin{tabular}[c]{@{}l@{}}$\alpha_{5},\dots,\alpha_{8},\alpha_{10},\alpha_{11},$ \\[-6pt] $\alpha_{12},\alpha_{14},\dotsc,\alpha_{19}=0$\end{tabular} & -- \\
$\beta_1,\dotsc,\beta_{14}=0$ & -- \\
$\gamma_1,\gamma_2,\gamma_3=0$ & -- \\[1pt]
\hline
\end{tabular}
\end{table}

\subsection{$\boldsymbol{7}$th-order low-density model}
The third and last reduced model that we present is the $7$th-order low-density model. It considers terms up to a total order in the density and derivatives of seven, i.e., it makes use of the low-density approximation. The density current of this model reads
\begin{equation}
\begin{split}
J_i &= (\alpha_{1}+\alpha_{2}\rho+\alpha_{3}\rho^2)\partial_i\rho+(\alpha_{4}+\alpha_{5}\rho+\alpha_{6}\rho^2)\partial_i\Laplace\rho\\
&\quad  +(\alpha_{9}+\alpha_{10}\rho)(\Laplace\rho)\partial_i\rho+(\alpha_{13}+\alpha_{14}\rho)(\partial_i\partial_j\rho)\partial_j\rho\\
&\quad +\alpha_{17}(\partial_j\rho)(\partial_j\rho)\partial_i\rho+\epsilon_1\partial_i\Laplace^2\rho
\end{split}\label{eqn:J7th}\raisetag{1.4em}
\end{equation}
with the coefficient 
\begin{equation}
\epsilon_1 = \dfrac{v_0^2}{48600 D_\mathrm{R}^3} \bigg(45D_\mathrm{T} 
+\dfrac{2 v_0^2}{D_\mathrm{R}}\bigg)^2-\dfrac{D_\mathrm{T}v_0^4}{1620 D_\mathrm{R}^4}
\end{equation}
and the constitutive equations for $\mathcal{P}_i$ and $\mathcal{Q}_{ij}$ are given by
\begin{equation}
\begin{split}
\mathcal{P}_i &= (\beta_{1}+\beta_{2}\rho)\partial_i\rho+(\beta_{3}+\beta_{4}\rho+\beta_{5}\rho^2)\partial_i\Laplace\rho \\
&\quad +(\beta_{7}+\beta_{8}\rho)(\Laplace\rho)\partial_i\rho+(\beta_{10}+\beta_{11}\rho)(\partial_i\partial_j\rho)\partial_j\rho\\
&\quad  +\beta_{13}(\partial_j\rho)(\partial_j\rho)\partial_i\rho-\dfrac{3 \epsilon_1}{4\pi v_0}\partial_i\Laplace^2\rho\label{eqn:P7th}
\end{split}\raisetag{2.2em}
\end{equation}
and 
\begin{equation}
\begin{split}
\mathcal{Q}_{ij} &= (\gamma_1+\gamma_2\rho)(3\partial_i\partial_j\rho+\delta_{ij}\Laplace\rho)\\
&\quad +\gamma_2(-3(\partial_i\rho)\partial_j\rho+\delta_{ij}(\partial_k\rho)\partial_k\rho),
\end{split}\label{eqn:7thOrderLDQ}%
\end{equation}
respectively. This model is a reduced $4$th-order-derivatives model with an additional term $\propto\partial_i\Laplace^2\rho$. Also this model has a similar form as the corresponding model for two spatial dimensions, which is given by Eqs.\ (20) and (28)-(31) in Ref.\ \cite{BickmannW2019}.

The term $\propto\partial_i\Laplace^2\rho$ with its fifth order in derivatives (leading to a contribution of sixth order in derivatives in the dynamic equation for $\rho$) is of particular importance, since it allows the $7$th-order low-density model to describe crystalline states on a particle-resolving length scale. Therefore, the model can also be called a \ZT{PFC model}. 
There exist already some \ZT{active PFC models} that describe crystals of ABPs \cite{Menzel2013,Menzel2014,Alaimo2016,Alaimo2018,PraetoriusVWL2018}, but they are inherently different from our model proposed here.
In contrast to our model, the active PFC models from the literature are applicable only close to thermodynamic equilibrium \cite{teVrugtLW2020}.  
Furthermore, they focus on particles in two spatial dimensions, where the two-dimensional manifold on which the particles move is planar in Refs.\ \cite{Menzel2013,Menzel2014,Alaimo2016,Alaimo2018} and spherical in Ref.\ \cite{PraetoriusVWL2018}. 
To the best of our knowledge, there presently exists no active PFC model for ABPs in three spatial dimensions. 
This means that our $7$th-order low-density model, given by Eqs.\ \eqref{eqn:Dyn2DwD} and \eqref{eqn:J7th}-\eqref{eqn:7thOrderLDQ}, constitutes a reasonable starting point for investigating the nonequilibrium dynamics of active crystals in three spatial dimensions, where one can expect to observe even more positional and orientational patterns than those that have been found for lower-dimensional active crystals \cite{Ophaus2018}.

\section{\label{sec:Applications}Applications}
The general field theory \eqref{eqn:formalDynEqOP} and already the simpler models presented in section \ref{sec:MODELS} have many possible applications. To present two examples, in the following we apply them to predict a spinodal condition describing the onset of MIPS and an expression for the density-dependent mean swimming speed of the interacting particles. 

\subsection{\label{ssec:MIPS}Predictions for motility-induced phase separation}
To obtain a spinodal condition that describes the onset of MIPS as a function of the P\'eclet number $\mathrm{Pe}$, which is a measure for the activity of the ABPs, and the particles' mean packing density $\Phi$ \cite{Bialk2013,Cates2013,Speck2014,WittkowskiSC2017,BickmannW2019}, we consider the $2$nd-order-derivatives model and perform a linear stability analysis of the dynamic equation for the concentration field \eqref{eqn:Dyn2DwD}. This yields the spinodal condition 
\begin{equation}
D(\rho) = 0,
\label{eqn:spinodal3D}%
\end{equation}
where $D(\rho)$ is given by Eq.\ \eqref{eqn:diffcoeffienetn}.
Similar as for the corresponding spinodal condition for two spatial dimensions, which is given by Eq.\ (66) in Ref.\ \cite{BickmannW2019}, there are three different coefficients, here given by $G(1, 0, 0, 0)$, $G(0, 1, 1, 0)$, and $G(0, 1, 0, 1)$, also in the spinodal condition for three spatial dimensions \eqref{eqn:spinodal3D}. 
Simulation results for MIPS of ABPs in three spatial dimensions can be found in Refs.\ \cite{Stenhammar2014, Wysocki2014, Broeker2019}. Reference \cite{Broeker2019} provides also an analytic representation of the pair-distribution function of ABPs that interact via the purely repulsive Weeks-Chandler-Andersen (WCA) potential, which is a standard choice for the pair-interaction potential of ABPs in simulations and well in line with the behavior of ABPs observed in experiments \cite{Buttinoni2013}. When denoting the diameter of the ABPs with $\sigma$ and introducing the interaction strength $\epsilon$, the WCA potential reads
\begin{equation}
U_2(r) = \begin{cases}
4\epsilon\Big((\frac{\sigma}{r})^{12}-(\frac{\sigma}{r})^{6}\Big)+\epsilon, &\text{if }r<2^{\frac{1}{6}}\sigma,\\
0, & \text{else}.
\end{cases}\label{eqn:WCA}%
\end{equation}
For the remainder of this work, we consider this interaction potential, which allows us to use the analytic representation of the pair-distribution function from Ref.\ \cite{Broeker2019}. The particular values of the three different coefficients of the spinodal condition \eqref{eqn:spinodal3D} are then approximately given by 
\begin{align}
G(1,0,0,0) &= 3.31 + 1.01 e^{4.07\Phi}, \label{eqn:GIOOO}\\
G(0,1,1,0) &= 1.79 + 0.561\Phi, \label{eqn:GOIIO}\\
G(0,1,0,1) &= -1.63, \label{eqn:GOIOI}
\end{align}
where $\Phi = \rho\sigma^3\pi/6$ is the mean packing density of the system. $G(1,0,0,0)$ follows an exponential function with a constant offset, which applies also to the corresponding coefficient in the model for two spatial dimensions \cite{BickmannW2019}. Interestingly, $G(0,1,1,0)$ and $G(0,1,0,1)$ are given by a linear function and a constant, whereas the corresponding coefficients in two spatial dimensions follow a constant and a linear function \cite{BickmannW2019}, respectively. This difference will be further addressed in section \ref{ssec:SwimSpeed}. 

\begin{figure}[htb]
\centering
\includegraphics{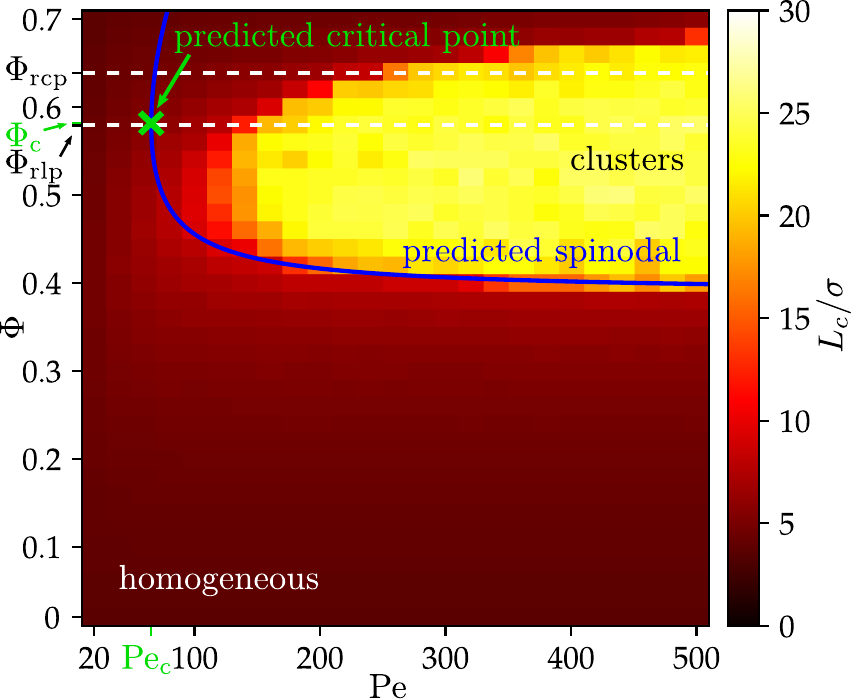}
\caption{\label{fig:MIPS}State diagram showing the characteristic length $L_c$ calculated from simulation data \cite{Broeker2019} as a function of the P\'{e}clet number $\mathrm{Pe}$ and packing density $\Phi$ of the system. Large values of $L_c$ correspond to clusters originating from MIPS and low values to a homogeneous state. Our predictions for the spinodal and associated critical point ($\textrm{Pe}_\textrm{c} = 67.9$, $\Phi_\textrm{c} = 0.584$) are also shown. Furthermore, the random-loose-packing density $\Phi_{\textrm{rlp}} = 0.58$ \cite{VISSCHER1972} and the random-close-packing density $\Phi_{\textrm{rcp}} = 0.64$ \cite{Li2010} for hard spheres in three spatial dimensions are indicated.}
\end{figure}

By using Eqs.\ \eqref{eqn:GIOOO}-\eqref{eqn:GOIOI} as well as the relations $D_\mathrm{T} = \sigma^2/\tau_{\mathrm{LJ}}$ with the Lennard-Jones time $\tau_{\mathrm{LJ}} = \sigma^2/(\beta D_{\mathrm{T}} \epsilon)$ and $D_\mathrm{R} = 3D_\mathrm{T}/\sigma^2$, which apply for spherical particles, it is possible to plot the spinodal condition \eqref{eqn:spinodal3D} as a function of the P\'eclet number $\textrm{Pe} = v_0\sigma/D_{\mathrm{T}}$ and mean packing density $\Phi$. The theoretical prediction and corresponding results of Brownian dynamics simulations from Ref.\ \cite{Broeker2019} are shown in Fig.\ \ref{fig:MIPS}.
In the simulations, the P\'eclet number was varied via $D_\mathrm{T}$, keeping the bare propulsion speed constant at $v_0 = 24\sigma/\tau_{\mathrm{LJ}}$, and the quantities $\sigma$, $\tau_{\mathrm{LJ}}$, and $\epsilon$ were used as units for length, time, and energy, respectively. 
A comparison of the analytic prediction and simulation data shows a good agreement for packing densities $\Phi < 0.6$. 
However, at higher packing densities $\Phi \gtrsim 0.6$ the agreement decreases more and more. A possible explanation is that at such high densities, which are larger than the random-loose-packing density $\Phi_{\textrm{rlp}} = 0.58$ \cite{VISSCHER1972} and close to or larger than the random-close-packing density $\Phi_{\textrm{rcp}} = 0.64$ \cite{Li2010} of hard spheres in three spatial dimensions, the characteristic length $L_c$ that is used in Fig.\ \ref{fig:MIPS} to identify MIPS is no longer suitable to distinguish clusters originating from MIPS from structure formation due to random clustering, percolation, and solidification phenomena. 
As a further result, we are able to give a prediction for the critical point. According to our equations, the coordinates of the critical point are given by $\textrm{Pe}_{\textrm{c}} = 67.9$ and $\Phi_{\textrm{c}} = 0.584$. 
To the best of our knowledge, there are no other analytic predictions for the spinodal or critical point of MIPS in systems of ABPs in three spatial dimensions in the literature with whom we could compare. 
Comparing instead with the available results for ABPs in two spatial dimensions, where the critical point is at ($\textrm{Pe}_{\textrm{c}, 2\textrm{D}} \approx 40, \Phi_{\textrm{c}, 2\textrm{D}} \approx 0.6$) \cite{Siebert2018,BickmannW2019}, one finds that MIPS requires a larger activity but no larger overall particle density in three spatial dimensions than in two spatial dimensions, which is in line with the findings described in Ref.\ \cite{Stenhammar2014}.

\subsection{\label{ssec:SwimSpeed}Density-dependent mean swimming speed}
\begin{figure}[htb]
\centering
\includegraphics{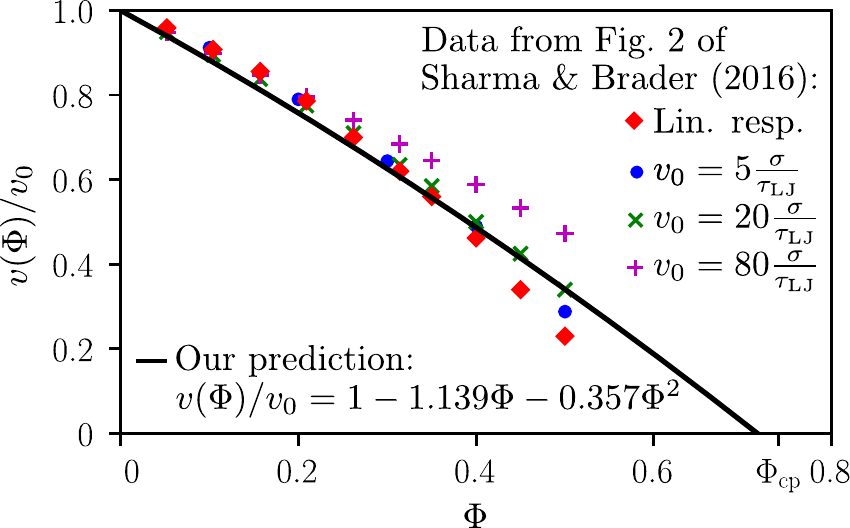}
\caption{\label{fig:3}Comparison of our prediction \eqref{eq:vPhi} for the mean swimming speed $v(\Phi)$ as a function of the particles' mean packing density $\Phi$ and corresponding results from Ref.\ \cite{Sharma2016} for the linear-response limit and three larger values of $v_0$. The close-packing density for hard spheres in three spatial dimensions is denoted by $\Phi_{\mathrm{cp}}=\pi/({3\sqrt{2}})\approx 0.74$.}
\end{figure}

The microscopic parameter $v_0$ denotes the bare propulsion speed of a single ABP or such particles in a very dilute system, where particle interactions can be neglected. If the interactions between the ABPs become important, the swimming speed reduces for increasing particle density $\rho$. 
The density dependence of the swimming speed $v[\rho]$ is an important quantity as it allows for conclusions on the particle interactions and it is known that a sufficiently fast decrease of the swimming speed with the local density can lead to the emergence of MIPS \cite{Tailleur2008,Cates2013}.
One can determine the density-dependent swimming speed $v[\rho]$ from Eq.\ \eqref{eqn:NeededforV} by performing an orientational expansion, a QSA, neglecting terms of second and higher orders in derivatives, and extracting the contribution to the dynamics of $\varrho$ of the form $\Nabla \cdot (v[\rho]\hat{u}\varrho(\vec{r}, \hat{u}, t))$. We find for the density-dependent swimming speed up to zeroth order in derivatives the expression
\begin{equation}
v(\rho) = v_0-\frac{2}{\pi}G(0, 110)\rho.
\label{eqn:SwimSpeedForm}%
\end{equation}
This result gives a predictive relation for the phenomenological parameter $v$ of the model for three spatial dimensions proposed in Ref.\ \cite{Cates2013}.
At a first glance, our result for the density-dependent swimming speed seems to decrease linearly with the density, as the result for $v(\rho)$ in two spatial dimensions from Ref.\ \cite{BickmannW2019} does. However, for the WCA interaction potential considered here, the coefficient $G(0, 110)$ has a dependence on $\rho$. Inserting the expression \eqref{eqn:GOIIO} into Eq.\ \eqref{eqn:SwimSpeedForm}, using the proportionality of $\rho$ and $\Phi$, and setting $v_0 = 24\sigma/\tau_{\mathrm{LJ}}$ (see Section \ref{ssec:MIPS}) yields
\begin{equation}
v(\Phi) \approx v_0(1-1.139\Phi-0.357\Phi^2).
\label{eq:vPhi}%
\end{equation}
This density dependence of the swimming speed is to first order a linear decrease, but interestingly we find a notable negative quadratic correction that becomes relevant for high densities. The same behavior was predicted in Ref.\ \cite{Sharma2016}, which is a work based on the Green-Kubo approach. Their numerical data are in excellent agreement with our prediction (see Fig.\ \ref{fig:3}). 

\section{\label{sec:Conclusion}Conclusions}
We derived a predictive local field theory for the collective dynamics of spherical ABPs in three spatial dimensions. The derived theory is rather general and highly accurate and includes order parameters and derivatives up to infinite orders. Alongside the general theory we provided less complex reduced models that can be favorable for specific applications. These special models contain various popular models from the literature, such as AMB \cite{WittkowskiTSAMC2014} and AMB+ \cite{Tjhung2018}, as limiting cases. Via a comparison we were able to obtain analytic expressions for the often phenomenological coefficients of the models from the literature, thus linking these coefficients to the microscopic parameters of the considered system.
The applicability and accuracy of our theory was demonstrated by addressing specific applications. We derived an analytic expression for the density-dependent mean swimming speed of ABPs in a homogeneous state and found an interesting high-density correction that seems to be of relevance solely for three-dimensional systems. This correction extends the usually considered linear density-dependence and is confirmed by results from the literature that were obtained by a different analytic approach \cite{Sharma2016}. Furthermore, we derived an analytic expression for the spinodal describing the onset of MIPS in a system of spherical ABPs in three spatial dimensions. A comparison of our spinodal condition with recent simulation results for particles interacting via a Weeks-Chandler-Andersen potential from the literature \cite{Broeker2019} showed a good agreement. To the best of our knowledge, this spinodal condition constitutes the first predictive analytic expression for the onset of MIPS in a system of ABPs with three spatial dimensions. Our spinodal condition also yielded a prediction for the critical point associated with the spinodal for which no other estimates seem to exist in the literature so far. In the future, detailed numerical studies of ABP systems in three spatial dimensions should give good estimates for the critical point, making it possible to confirm our prediction.

The general field theory and simpler reduced models presented in this work can be applied to study a broad range of further far-from-equilibrium effects in systems of ABPs. With the $4$th-order-derivatives model one could, e.g., investigate the clustering and formation of interfaces in the dynamics of ABPs at high densities, which is not fully captured by AMB and AMB+ as they do not include all potentially relevant terms with a high order in the density field. The $7$th-order low-density model could be used to study effects like solidification and crystallization in systems of ABPs. This model can be seen as an extension of the traditional active PFC model \cite{Menzel2013} towards far-from-equilibrium contributions and three spatial dimensions. One can expect that many fascinating findings are still to discover in solid states of ABPs, especially in three spatial dimensions.
As in the case of two spatial dimensions \cite{BickmannW2019}, an application of our reduced models to systems of ABPs on curved manifolds like a sphere \cite{PraetoriusVWL2018} is straightforwardly possible, since the density currents in these models contain only a scalar order-parameter field where the occurring differential operators can easily be adapted to a curved manifold. 
If needed, also more complex and accurate models can be derived from our general field theory by omitting some of the approximations made for simplification. 
Finally, one could extend our general field theory towards even more complex systems. Important examples of such systems are mixtures of different types of active and passive particles \cite{StenhammarWMC2015,WittkowskiSC2017,Alaimo2018} and active particles with nonspherical shapes \cite{WittkowskiL2011,WittkowskiL2012} as in active liquid crystals \cite{DeCamp2015,Doostmohammadi2018,Lemma2019}. Since controlling the dynamics of active colloidal particles is a highly relevant topic when envisaging applications of active particles in fields like medicine and materials science, an extension of the field theory towards the inclusion of external fields will be a further important quest for the future.

\begin{acknowledgments}
We thank Andreas Menzel and Uwe Thiele for helpful discussions and Joseph Brader and Abhinav Sharma for providing the comparative data for Fig.\ \ref{fig:3}. 
R.W.\ is funded by the Deutsche Forschungsgemeinschaft (DFG, German Research Foundation) -- WI 4170/3-1. 
\end{acknowledgments}

\appendix
\section{Coefficients for the $\boldsymbol{4}$th-order-derivatives model}\label{sec:Koeff4th}
In this appendix, we give explicit expressions for the coefficients that occur in Eqs.\ \eqref{eqn:4thorderCurrentJ}-\eqref{eqn:7thOrderLDQ}. 
For simplicity, the rotational relaxation time $\tau = 1/D_\mathrm{R}$ is introduced.

The coefficients occurring in Eqs.\ \eqref{eqn:4thorderCurrentJ} and \eqref{eqn:J7th} are given by
\begin{align}
\alpha_{1} &= D_\mathrm{T}+\frac{1}{6}\tau^2v_0^2,\\
\begin{split}
    \alpha_{2} &= -\frac{1}{3\pi}\tau v_0(G(0,1,0,1)+3G(0,1,1,0))\\
    &\quad\, +\frac{2}{3\pi}G(1,0,0,0),
\end{split}\\
\alpha_{3} &= \frac{4}{3\pi^2}\tau G(0,1,1,0)(G(0,1,0,1)+G(0,1,1,0)),\\
\alpha_{4} &= \frac{1}{540} \tau ^2 v_0^2 (45 D_\mathrm{T}+2 \tau  v_0^2),\\
\begin{split}
    \alpha_{5} &= \frac{1}{270 \pi}(\tau  v_0(-45 D_\mathrm{T} \tau  (G(0,1,0,1)\\
    &\quad\,+3 G(0,1,1,0))+\tau  v_0(\tau  v_0 (-2 G(0,1,0,1)\\
    &\quad\,-10 G(0,1,1,0)+\sqrt{2} (G(0,1,1,2)\\
    &\quad\,+G(0,1,2,1)))-5 \sqrt{3} G(1,0,1,1)\\
    &\quad\,+4 G(1,2,0,2)+2 \sqrt{30} G(1,2,1,1)\\
    &\quad\,+4 G(1,2,2,0))-27 (G(2,1,0,1)\\
    &\quad\,+G(2,1,1,0)))+18 G(3,0,0,0)),
\end{split}\\
\begin{split}
    \alpha_{6} &= \frac{1}{135\pi^2} \tau (90 D_\mathrm{T} \tau  G(0,1,1,0) (G(0,1,0,1)\\
    &\quad\,+G(0,1,1,0))+\tau ^2 v_0^2(18 G(0,1,1,0)^2\\
    &\quad\,+G(0,1,1,2) G(0,1,2,1)-4 \sqrt{2} (G(0,1,1,2)\\
    &\quad\,+G(0,1,2,1)) G(0,1,1,0)+G(0,1,0,1) \\
    &\quad\quad\,\, (8 G(0,1,1,0)-\sqrt{2} (G(0,1,1,2)\\
    &\quad\,+G(0,1,2,1))))+\tau  v_0 (G(0,1,1,0) \\
    &\quad\,\quad\,(15 \sqrt{3} G(1,0,1,1)-12 G(1,2,0,2)\\
    &\quad\,-6 \sqrt{30} G(1,2,1,1)-8 G(1,2,2,0))\\
    &\quad\,+G(0,1,0,1) (5 \sqrt{3} G(1,0,1,1)\\
    &\quad\,-2 \sqrt{30} G(1,2,1,1)-4 G(1,2,2,0))\\
    &\quad\, +2 \sqrt{2} (G(0,1,2,1) G(1,2,0,2)\\
    &\quad\,+G(0,1,1,2) G(1,2,2,0)))+8 G(1,2,0,2) \\
    &\quad\,\quad\,G(1,2,2,0)+54 G(0,1,1,0) G(2,1,0,1)\\
    &\quad\,+27 (G(0,1,0,1)+G(0,1,1,0)) G(2,1,1,0)),
\end{split}\raisetag{12em}\\
\begin{split}
    \alpha_{7} &= \frac{2}{135\pi^3} \tau ^2 (-14 \tau  v_0 G(0,1,1,0)^3\\
    &\quad\,+G(0,1,1,0)^2(5 \tau  v_0 (\sqrt{2}G(0,1,1,2)\\
    &\quad\,+\sqrt{2}G(0,1,2,1)-2 G(0,1,0,1))\\
    &\quad\,-10 \sqrt{3} G(1,0,1,1)+8G(1,2,0,2)\\
    &\quad\,+4\sqrt{30}G(1,2,1,1)+4G(1,2,2,0)) \\
    &\quad\,+G(0,1,1,0)( 3 \tau  v_0(\sqrt{2} G(0,1,0,1) \\
    &\quad\,\quad\,(G(0,1,1,2)+G(0,1,2,1)) \\
    &\quad\, -G(0,1,1,2) G(0,1,2,1))\\
    &\quad\,-2 \sqrt{2} (2 G(0,1,2,1) G(1,2,0,2)\\
    &\quad\,+G(0,1,1,2) G(1,2,2,0))\\
    &\quad\,+G(0,1,0,1) (4\sqrt{30}G(1,2,1,1)\\
    &\quad\,+4G(1,2,2,0)-10 \sqrt{3} G(1,0,1,1)))\\
    &\quad\, -G(0,1,0,1) G(0,1,1,2)(\tau  v_0 G(0,1,2,1) \\
    &\quad\, +2 \sqrt{2} G(1,2,2,0))),
\end{split}\raisetag{25ex}\\
\begin{split}
    \alpha_{8} &= \frac{8}{135\pi^4} \tau ^3 G(0,1,1,0) (G(0,1,0,1)\\
    &\quad\,+G(0,1,1,0))(-\sqrt{2}G(0,1,1,0)\\
    &\quad\,\quad\,(G(0,1,1,2)+G(0,1,2,1)) \\
    &\quad\,+2 G(0,1,1,0)^2 +G(0,1,1,2) G(0,1,2,1)),
\end{split}\raisetag{8ex}\\
\begin{split}
    \alpha_{9} &= -\frac{1}{1080\pi} \tau v_0 (20 \tau (18 D_\mathrm{T}+\tau v_0^2) G(0,1,1,0)\\
    &\quad\,+\tau  v_0 (\tau v_0 (30 G(0,1,0,1)+4\sqrt{2}G(0,1,1,2)\\
    &\quad\,-5\sqrt{2} G(0,1,2,1))-60 G(1,0,0,0)\\
    &\quad\,-12\sqrt{30} G(1,2,1,1)-20 G(1,2,2,0))\\
    &\quad\,+72 G(2,1,1,0)),
\end{split}\raisetag{4em}\\
\begin{split}
    \alpha_{10} &= \frac{1}{540\pi^2} \tau (72 G(0,1,1,0) (5 D_\mathrm{T} \tau  (3 G(2,1,0,1)\\
    &\quad\,+G(0,1,0,1)+G(0,1,1,0)))\\
    &\quad\,+\tau ^2 v_0^2( G(0,1,0,1)( (2\sqrt{2} G(0,1,1,2)\\
    &\quad\,-7\sqrt{2} G(0,1,2,1))+170 G(0,1,1,0))\\
    &\quad\,+30 G(0,1,0,1)^2+64 G(0,1,1,0)^2\\
    &\quad\, -2 G(0,1,2,1)^2 +3 \sqrt{2} G(0,1,1,0) (2 G(0,1,1,2)\\
    &\quad\,-9 G(0,1,2,1)))-2 \tau  v_0(4 \sqrt{2} G(0,1,2,1) \\
    &\quad\,\quad\, (G(1,2,2,0)-G(1,2,0,2))+G(0,1,0,1) \\
    &\quad\,\quad\, (30 G(1,0,0,0)+3 \sqrt{30} G(1,2,1,1)\\
    &\quad\,+14 G(1,2,2,0))+G(0,1,1,0) (90 G(1,0,0,0)\\
    &\quad\,-20 \sqrt{3} G(1,0,1,1)+29 \sqrt{30} G(1,2,1,1)\\
    &\quad\,+24 G(1,2,0,2)+18 G(1,2,2,0))) \\
    &\quad\,+16 G(1,2,2,0)(2 G(1,2,0,2)-G(1,2,2,0))\\
    &\quad\,+36 G(2,1,1,0)(G(0,1,0,1)+3 G(0,1,1,0)) ),
\end{split}\raisetag{11.6em}\\
\begin{split}
    \alpha_{11} &= \frac{1}{135\pi^3} \tau ^2(G(0,1,1,0)^2(\tau  v_0 ( (\sqrt{2}G(0,1,1,2)\\
    &\quad\,+19\sqrt{2} G(0,1,2,1))-142 G(0,1,0,1))\\
    &\quad\,+60 G(1,0,0,0)-20 \sqrt{3} G(1,0,1,1)\\
    &\quad\,+32 G(1,2,0,2)+26 \sqrt{30} G(1,2,1,1)\\
    &\quad\,+8 G(1,2,2,0))-34 \tau  v_0 G(0,1,1,0)^3\\
    &\quad\,+G(0,1,1,0)(\tau  v_0 (-60 G(0,1,0,1)^2\\
    &\quad\,+3 \sqrt{2} G(0,1,0,1) (G(0,1,1,2) \\
    &\quad\, +5 G(0,1,2,1))+G(0,1,2,1) (4 G(0,1,2,1)\\
    &\quad\,-3 G(0,1,1,2)))+2 (G(0,1,0,1)\\
    &\quad\,\quad\, (30 G(1,0,0,0)-10 \sqrt{3} G(1,0,1,1)\\
    &\quad\,+7 \sqrt{30} G(1,2,1,1)+8 G(1,2,2,0))\\
    &\quad\, + \sqrt{2}G(1,2,2,0)(G(0,1,1,2)+4 G(0,1,2,1)) \\
    &\quad\, -8\sqrt{2} G(0,1,2,1) G(1,2,0,2)) ) \\
    &\quad\, -G(0,1,0,1) G(0,1,1,2)\\
    &\quad\, \quad\, (\tau  v_0 G(0,1,2,1)+2 \sqrt{2} G(1,2,2,0))),
\end{split}\raisetag{14em}\\
\begin{split}
    \alpha_{12} &= \frac{4}{135\pi^4} \tau ^3 G(0,1,1,0) ( G(0,1,1,0)^2\\
    &\quad\,\quad\,(36 G(0,1,0,1)-\sqrt{2} (G(0,1,1,2)\\
    &\quad\, +4 G(0,1,2,1))) +6 G(0,1,1,0)^3\\
    &\quad\,+G(0,1,1,0)(-3 \sqrt{2} G(0,1,0,1)(G(0,1,1,2) \\
    &\quad\,+2 G(0,1,2,1)) +30 G(0,1,0,1)^2\\
    &\quad\, +G(0,1,2,1) (G(0,1,1,2)-2 G(0,1,2,1) ) )\\
    &\quad\, +3 G(0,1,0,1) G(0,1,1,2) G(0,1,2,1)),
\end{split}\raisetag{7em}\\
\begin{split}
    \alpha_{13} &= \frac{1}{1080\pi} \tau  v_0 (-60 \tau (12 D_\mathrm{T}+\tau v_0^2) G(0,1,1,0)\\
    &\quad\, +\tau  v_0 (\tau v_0 (12\sqrt{2} G(0,1,1,2)\\
    &\quad\,+35\sqrt{2} G(0,1,2,1)-30 G(0,1,0,1))\\
    &\quad\,+60 G(1,0,0,0)-40 \sqrt{3} G(1,0,1,1)\\
    &\quad\,+4\sqrt{30} G(1,2,1,1)+140 G(1,2,2,0))\\
    &\quad\, -144 G(2,1,1,0)),
\end{split}\raisetag{4em}\\
\begin{split}
    \alpha_{14} &= \frac{1}{540\pi^2} \tau (720 D_\mathrm{T} \tau G(0,1,1,0) (G(0,1,0,1)\\
    &\quad\,+G(0,1,1,0))+48 G(1,2,2,0)^2\\
    &\quad\,+64 G(1,2,0,2) G(1,2,2,0)\\
    &\quad\,+432 G(0,1,1,0) G(2,1,0,1)\\
    &\quad\,+72 G(0,1,0,1) G(2,1,1,0)+216 G(0,1,1,0)\\
    &\quad\,\quad\, G(2,1,1,0)+\tau^2 v_0^2(30 G(0,1,0,1)^2\\
    &\quad\,+G(0,1,0,1)(210 G(0,1,1,0)\\
    &\quad\,-6\sqrt{2} G(0,1,1,2)-29 \sqrt{2} G(0,1,2,1)) \\
    &\quad\,+208 G(0,1,1,0)^2+2 G(0,1,2,1)\\
    &\quad\,\quad\,(10 G(0,1,1,2)+3 G(0,1,2,1)) \\
    &\quad\,-3 \sqrt{2} G(0,1,1,0) (26 G(0,1,1,2) \\
    &\quad\,+63 G(0,1,2,1)))\\
    &\quad\, -2 \tau v_0 (G(0,1,0,1)(30 G(1,0,0,0)\\
    &\quad\,-10 \sqrt{3} G(1,0,1,1)+\sqrt{30} G(1,2,1,1)\\
    &\quad\,+58 G(1,2,2,0))+G(0,1,1,0)\\
    &\quad\,\quad\, (90 G(1,0,0,0)+48 G(1,2,0,2)\\
    &\quad\,-110 \sqrt{3} G(1,0,1,1)+23 \sqrt{30} G(1,2,1,1)\\
    &\quad\,- 20 \sqrt{2} G(0,1,1,2)G(1,2,2,0)\\
    &\quad\, -4 \sqrt{2}G(0,1,2,1) (2 G(1,2,0,2)\\
    &\quad\, +3 G(1,2,2,0))+206 G(1,2,2,0)) )),
\end{split}\raisetag{16em}\\
\begin{split}
    \alpha_{15} &= -\frac{1}{135\pi^3} \tau ^2 (\tau  v_0 (60 G(0,1,1,0) G(0,1,0,1)^2\\
    &\quad\,+G(0,1,0,1)(194 G(0,1,1,0)^2\\
    &\quad\,-\sqrt{2}G(0,1,1,0) (21 G(0,1,1,2)\\
    &\quad\,+65 G(0,1,2,1)) +7 G(0,1,1,2) G(0,1,2,1)) \\
    &\quad\,+G(0,1,1,0) (118 G(0,1,1,0)^2\\
    &\quad\, -\sqrt{2} G(0,1,1,0) (67 G(0,1,1,2) \\
    &\quad\, +153 G(0,1,2,1)) +G(0,1,2,1)\\
    &\quad\, \quad\, (41 G(0,1,1,2)+12 G(0,1,2,1))))\\
    &\quad\, -2G(0,1,0,1)(G(0,1,1,0)(30 G(1,0,0,0)\\
    &\quad\,-30 \sqrt{3} G(1,0,1,1)+9 \sqrt{30} G(1,2,1,1)\\
    &\quad\,+56 G(1,2,2,0)) -7 \sqrt{2} G(0,1,1,2) G(1,2,2,0))\\
    &\quad\, - 2G(0,1,1,0)(30 G(1,0,0,0)\\
    &\quad\, +32 G(1,2,0,2)-50 \sqrt{3} G(1,0,1,1)\\
    &\quad\,+11 \sqrt{30} G(1,2,1,1)+68 G(1,2,2,0)) \\
    &\quad\, +2\sqrt{2} G(0,1,1,0)(13G(0,1,1,2)\\
    &\quad\,\quad\, G(1,2,2,0)+4 G(0,1,2,1) \\
    &\qquad\,\, (4 G(1,2,0,2)+3 G(1,2,2,0)) ) ),
\end{split}\raisetag{14em}\\
\begin{split}
    \alpha_{16} &= \frac{4}{135\pi^4} \tau ^3 G(0,1,1,0)(30 G(0,1,1,0) G(0,1,0,1)^2\\
    &\quad\,+G(0,1,0,1)(-\sqrt{2} G(0,1,1,0) \\
    &\quad\,\quad\,(11 G(0,1,1,2)+32 G(0,1,2,1))\\
    &\quad\,+52 G(0,1,1,0)^2 +11 G(0,1,1,2) \\
    &\quad\,\quad\,G(0,1,2,1))+G(0,1,1,0)(22 G(0,1,1,0)^2\\
    &\quad\, -\sqrt{2}  G(0,1,1,0)(17 G(0,1,1,2)\\
    &\quad\,+38 G(0,1,2,1))+G(0,1,2,1) \\
    &\quad\,\quad\,(17 G(0,1,1,2)+6 G(0,1,2,1)))),
\end{split}\raisetag{7em}\\
\begin{split}
    \alpha_{17} &= \frac{1}{135\pi^2} \tau  (\tau ^2 v_0^2 (14 G(0,1,1,0)^2+2 G(0,1,1,0)\\
    &\quad\,\quad\, (15 G(0,1,0,1) -3\sqrt{2} G(0,1,1,2)\\
    &\quad\,-10\sqrt{2} G(0,1,2,1)) +G(0,1,2,1) (2 G(0,1,1,2)\\
    &\quad\,+G(0,1,2,1)))+ 2 \tau  v_0(2 \sqrt{2} G(1,2,2,0)\\
    &\quad\,\quad\,(G(0,1,1,2)+G(0,1,2,1)) \\
    &\quad\,-G(0,1,1,0)(15 G(1,0,0,0)\\
    &\quad\,-10 \sqrt{3} G(1,0,1,1)+4 \sqrt{30} G(1,2,1,1)\\
    &\quad\,+20 G(1,2,2,0)))+8 G(1,2,2,0)^2),
\end{split}\raisetag{6.5em}\\
\begin{split}
    \alpha_{18} &= \frac{2}{135\pi^3} \tau ^2(-30 \tau  v_0 G(0,1,1,0)^3+G(0,1,1,0)^2\\
    &\qquad\,\, (\tau  v_0 (21 \sqrt{2} G(0,1,1,2) +59 \sqrt{2} G(0,1,2,1)\\
    &\quad\,-104 G(0,1,0,1)) +30 G(1,0,0,0)\\
    &\quad\,-30 \sqrt{3} G(1,0,1,1) +12 \sqrt{30} G(1,2,1,1)\\
    &\quad\,+16 G(1,2,0,2) +44 G(1,2,2,0))\\
    &\quad\, -G(0,1,1,0)(\tau  v_0 (30 G(0,1,0,1)^2\\
    &\quad\,-\sqrt{2} G(0,1,0,1)(3 G(0,1,1,2) \\
    &\quad\,+17 G(0,1,2,1))+3 G(0,1,2,1) \\
    &\quad\, \quad\, (5 G(0,1,1,2)+2 G(0,1,2,1)))\\
    &\quad\, -2 G(0,1,0,1)(-5 \sqrt{3} G(1,0,1,1)\\
    &\quad\,+15 G(1,0,0,0)+2 \sqrt{30} G(1,2,1,1)\\
    &\quad\, +14 G(1,2,2,0))+ 8\sqrt{2} G(0,1,2,1) \\
    &\quad\,\quad\,G(1,2,0,2)+6\sqrt{2} G(1,2,2,0)(G(0,1,1,2)\\
    &\quad\, +2 G(0,1,2,1)) ) -G(0,1,0,1)G(0,1,1,2) \\
    &\quad\,\quad\, (\tau  v_0 G(0,1,2,1) +2 \sqrt{2} G(1,2,2,0))),
\end{split}\raisetag{12.6em}\\
\begin{split}
    \alpha_{19} &= \frac{16}{135\pi^4} \tau ^3 G(0,1,1,0) (G(0,1,1,0)^2\\
    &\quad\,\quad\,(19 G(0,1,0,1)-2 \sqrt{2} (2 G(0,1,1,2)\\
    &\quad\,+5 G(0,1,2,1)))+4 G(0,1,1,0)^3 \\
    &\quad\,+G(0,1,1,0)(15 G(0,1,0,1)^2\\
    &\quad\,-2 \sqrt{2} G(0,1,0,1)(G(0,1,1,2) \\
    &\quad\,+4 G(0,1,2,1))+2 G(0,1,2,1)  \\
    &\quad\,\quad\,(2 G(0,1,1,2)+G(0,1,2,1)))\\
    &\quad\, +2 G(0,1,0,1) G(0,1,1,2) G(0,1,2,1)).
\end{split}\raisetag{7em}
\end{align}
The coefficients occurring in Eqs.\ \eqref{eqn:P4th} and \eqref{eqn:P7th} are given by
\begin{align}
\beta_{1} &= -\frac{1}{8\pi}\tau v_0,\\
\beta_{2} &= \frac{1}{2\pi^2}\tau G(0,1,1,0),\\
\begin{split}
    \beta_{3} &= -\frac{1}{720\pi} \tau ^2 v_0 (45 D_\mathrm{T}+2 \tau  v_0^2),
\end{split}\\
\begin{split}
    \beta_{4} &= \frac{1}{360\pi^2} \tau (\tau  v_0 (\tau  v_0 (8 G(0,1,1,0) -\sqrt{2}G(0,1,1,2)\\
    &\quad\, -\sqrt{2}G(0,1,2,1)) -2 \sqrt{30} G(1,2,1,1)\\
    &\quad\, +5 \sqrt{3} G(1,0,1,1)-4 G(1,2,2,0))\\
    &\quad\,+90 D_\mathrm{T} \tau  G(0,1,1,0)+27 G(2,1,1,0)),
\end{split}\raisetag{3.5em}\\
\begin{split}
    \beta_{5} &= \frac{1}{180\pi^3} \tau ^2( G(0,1,1,0)(3 \sqrt{2} \tau  v_0 (G(0,1,1,2)\\
    &\quad\, +G(0,1,2,1))-10 \sqrt{3} G(1,0,1,1) \\
    &\quad\, +4G(1,2,2,0)+4\sqrt{30} G(1,2,1,1))  \\
    &\quad\,-G(0,1,1,2)(\tau  v_0 G(0,1,2,1)\\
    &\quad\, +2 \sqrt{2} G(1,2,2,0))-10 \tau  v_0 G(0,1,1,0)^2),
\end{split}\\
\begin{split}
    \beta_{6} &= \frac{1}{45\pi^4} \tau ^3 G(0,1,1,0)(2 G(0,1,1,0)^2\\
    &\quad\, -\sqrt{2} G(0,1,1,0)(G(0,1,1,2)\\
    &\quad\, +G(0,1,2,1))+G(0,1,1,2) G(0,1,2,1)),
\end{split}\\
\begin{split}
    \beta_{7} &= \frac{1}{1440\pi^2} \tau  (20 \tau (18 D_\mathrm{T}+\tau v_0^2) G(0,1,1,0)\\
    &\quad\, +\tau  v_0 (\tau  v_0(2\sqrt{2}  G(0,1,1,2)-7\sqrt{2}  G(0,1,2,1)\\
    &\quad\, + 30 G(0,1,0,1)) -6 \sqrt{30} G(1,2,1,1)\\
    &\quad -60 G(1,0,0,0)-28 G(1,2,2,0))\\
    &\quad\,+36 G(2,1,1,0)),
\end{split}\raisetag{4em}\\
\begin{split}
    \beta_{8} &= \frac{1}{360\pi^3} \tau ^2 (G(0,1,1,0)(3 \tau  v_0 (\sqrt{2} G(0,1,1,2)\\
    &\quad\, +5\sqrt{2}  G(0,1,2,1)-20 G(0,1,0,1))\\
    &\quad\, +60 G(1,0,0,0)-20 \sqrt{3} G(1,0,1,1)\\
    &\quad\, +14 \sqrt{30} G(1,2,1,1)+16 G(1,2,2,0))\\
    &\quad\,  -G(0,1,1,2)(\tau  v_0 G(0,1,2,1)\\
    &\quad\, +2 \sqrt{2} G(1,2,2,0))-22 \tau  v_0 G(0,1,1,0)^2),
\end{split}\\
\begin{split}
    \beta_{9} &= \frac{1}{30\pi^4} \tau ^3 G(0,1,1,0)(G(0,1,1,0)(10 G(0,1,0,1)\\
    &\quad\,-\sqrt{2} G(0,1,1,2)-\sqrt{2} 2 G(0,1,2,1)) \\
    &\quad\, +2 G(0,1,1,0)^2 +G(0,1,1,2) G(0,1,2,1)),
\end{split}\raisetag{2.8em}\\
\begin{split}
    \beta_{10} &= \frac{1}{1440\pi^2} \tau (60 \tau (12 D_\mathrm{T}+\tau  v_0^2) G(0,1,1,0) +\tau  v_0 \\
    &\quad\, \quad\, (\tau  v_0(30 G(0,1,0,1)-29\sqrt{2} G(0,1,2,1))\\
    &\quad\, -6\sqrt{2} G(0,1,1,2) +20 \sqrt{3} G(1,0,1,1)\\
    &\quad\, -60 G(1,0,0,0)-2\sqrt{30} G(1,2,1,1)\\
    &\quad\, -116 G(1,2,2,0))+72 G(2,1,1,0)),
\end{split}\raisetag{3.4em}\\
\begin{split}
    \beta_{11} &= \frac{1}{360\pi^3} \tau ^2 (G(0,1,1,0)(\tau  v_0 ( 21\sqrt{2} G(0,1,1,2)\\
    &\quad\, +65\sqrt{2} G(0,1,2,1)-60 G(0,1,0,1))\\
    &\quad\, -60 \sqrt{3} G(1,0,1,1)+18 \sqrt{30} G(1,2,1,1)\\
    &\quad\, +60 G(1,0,0,0)+112 G(1,2,2,0))\\
    &\quad\, -74 \tau  v_0 G(0,1,1,0)^2-7 G(0,1,1,2) \\
    &\quad\,\quad\, (\tau  v_0 G(0,1,2,1)+2 \sqrt{2} G(1,2,2,0))),
\end{split}\\
\begin{split}
    \beta_{12} &= \frac{1}{90\pi^4} \tau ^3 G(0,1,1,0) (G(0,1,1,0)\\
    &\qquad\, \, (-11\sqrt{2} G(0,1,1,2)-32\sqrt{2} G(0,1,2,1)\\
    &\quad\, +30 G(0,1,0,1)) +22 G(0,1,1,0)^2\\
    &\quad\, +11 G(0,1,1,2) G(0,1,2,1)),
\end{split}\\
\begin{split}
    \beta_{13} &= \frac{1}{180\pi^3} \tau ^2 ( G(0,1,1,0)(\tau  v_0 ( 3\sqrt{2} G(0,1,1,2)\\
    &\quad\, +17\sqrt{2} G(0,1,2,1)-30 G(0,1,0,1))\\
    &\quad\, +30 G(1,0,0,0)-10 \sqrt{3} G(1,0,1,1)\\
    &\quad+4 \sqrt{30} G(1,2,1,1)+28 G(1,2,2,0))\\
    &\quad\, -14 \tau  v_0 G(0,1,1,0)^2-G(0,1,1,2) \\
    &\quad\,\quad\, (\tau  v_0 G(0,1,2,1)+2 \sqrt{2} G(1,2,2,0))),
\end{split}\\
\begin{split}
    \beta_{14} &= \frac{2}{45\pi^4} \tau ^3 G(0,1,1,0) (G(0,1,1,0)(15 G(0,1,0,1)\\
    &\quad\, -2 \sqrt{2}G(0,1,1,2)-8 \sqrt{2} G(0,1,2,1)) \\
    &\quad\,+4 G(0,1,1,0)^2+2 G(0,1,1,2) G(0,1,2,1)).
\end{split}\raisetag{3em}
\end{align}
The coefficients occurring in Eqs.\ \eqref{eqn:Q4th} and \eqref{eqn:7thOrderLDQ} are given by
\begin{align}
\begin{split}
    \gamma_1 &= \frac{1}{144\pi} \tau ^2 v_0^2,
\end{split}\\
\begin{split}
    \gamma_2 &= -\frac{1}{144\pi^2} \tau  (\tau  v_0 (6 G(0,1,1,0)-\sqrt{2} G(0,1,2,1))\\
    &\quad\, -4 G(1,2,2,0)),
\end{split}\raisetag{2em}\\
\begin{split}
    \gamma_3 &= \frac{1}{36\pi^3} \tau ^2 G(0,1,1,0) (2 G(0,1,1,0) -\sqrt{2} G(0,1,2,1)).
\end{split}
\end{align}

\clearpage
\nocite{apsrev41Control}
\bibliographystyle{apsrev4-1}
\bibliography{control,refs}

\end{document}